\documentclass[aps,prc,twocolumn,showpacs,superscriptaddress,floatfix]{revtex4}
\usepackage{graphicx}

\begin{document}

\title{One-neutron knockout from $^{57}$Ni}

\author{K. L.~Yurkewicz}
 \altaffiliation[Present Address: ]{Fermi National Accelerator
 Laboratory, Batavia, IL 60510-0500} 
\affiliation{National Superconducting Cyclotron Laboratory, Michigan
 State University, East Lansing, MI 48824}
\affiliation{Department of Physics and Astronomy, Michigan State
 University, East Lansing, MI 48824} 
\author{D.~Bazin}  
\affiliation{National Superconducting Cyclotron Laboratory, Michigan
 State University, East Lansing, MI 48824} 
\author{B. A.~Brown}
\affiliation{National Superconducting Cyclotron Laboratory, Michigan
 State University, East Lansing, MI 48824} 
\affiliation{Department of Physics and Astronomy, Michigan State
 University, East Lansing, MI 48824} 
\author{J.~Enders} 
 \altaffiliation[Present Address: ]{Institut f{\"u}r Kernphysik,
 Technische Universit{\"a}t Darmstadt, Darmstadt, Germany} 
\affiliation{National Superconducting Cyclotron Laboratory, Michigan
 State University, East Lansing, MI 48824} 
\author{A.~Gade}
\affiliation{National Superconducting Cyclotron Laboratory, Michigan
 State University, East Lansing, MI 48824} 
\affiliation{Department of Physics and Astronomy, Michigan State
 University, East Lansing, MI 48824} 
\author{T.~Glasmacher}
 \email[E-mail: ]{glasmacher@nscl.msu.edu}
\affiliation{National Superconducting Cyclotron Laboratory, Michigan
 State University, East Lansing, MI 48824} 
\affiliation{Department of Physics and Astronomy, Michigan State
 University, East Lansing, MI 48824} 
\author{P. G.~Hansen}
\affiliation{National Superconducting Cyclotron Laboratory, Michigan
 State University, East Lansing, MI 48824} 
\affiliation{Department of Physics and Astronomy, Michigan State
 University, East Lansing, MI 48824} 
\author{V.~Maddalena}
 \altaffiliation[Present Address: ]{Universita di Basilicata, Italy}
\affiliation{National Superconducting Cyclotron Laboratory, Michigan
 State University, East Lansing, MI 48824} 
\affiliation{Department of Physics and Astronomy, Michigan State
 University, East Lansing, MI 48824} 
\author{A.~Navin}
 \altaffiliation[Present Address: ]{GANIL, Boite Postale 5027,
 F-14076, Caen Cedex, France} 
\affiliation{National Superconducting Cyclotron Laboratory, Michigan
 State University, East Lansing, MI 48824} 
\author{B. M.~Sherrill}
\affiliation{National Superconducting Cyclotron Laboratory, Michigan
 State University, East Lansing, MI 48824} 
\affiliation{Department of Physics and Astronomy, Michigan State
 University, East Lansing, MI 48824} 
\author{J.A.~Tostevin}
\affiliation{Department of Physics, School of Electronics and Physical
 Sciences, University of Surrey, Guildford, Surrey GU2 7XH, United
 Kingdom} 

\date{\today}

\begin{abstract}
The single-particle structure of $^{57}$Ni and level structure of
$^{56}$Ni were investigated with the \mbox{$^{9}$Be ($^{57}
$Ni,$^{56}$Ni+$\gamma$)$\it{X}$} reaction at 73~MeV/nucleon. An inclusive
cross section of 41.4(12)~mb was obtained for the reaction,
compared to a theoretical prediction of 85.4 mb, hence only 48(2)\% of the
theoretical cross section is exhausted.  This reduction in the observed
spectroscopic strength is consistent with that found for lighter
well-bound nuclei.  One-neutron removal spectroscopic factors of
0.58(11) to the ground state and 3.7(2) to all excited states of $^{56}$Ni
were deduced.
\end{abstract}

\pacs{21.10.Jx; 27.40.+z; 25.60.Gc}

\maketitle

\section{Introduction}

Doubly-magic nuclei are key benchmarks to our understanding of nuclear
shell structure. Until experiments around $^{100}$Sn become feasible, 
$^{56}$Ni is the heaviest experimentally accessible doubly-magic
nucleus in which protons  
and neutrons occupy the same orbitals.
An early neutron-pickup experiment populating
states in  $^{57}$Ni suggested that the low-lying excited states
may not be purely single-particle in nature, but may include
contributions from $^{56}$Ni excited states \cite{goul69}. The
$\beta$-decay of $^{57}$Cu  
indicated \cite{Shi84} that the Gamov-Teller matrix element of its
decay to the ground state  
of $^{57}$Ni is significantly reduced compared to the single-particle
value. 
Measurements of unusually high collectivity of $^{56}$Ni
\cite{krau94,yani98,yur04} reinforced 
the questions about the single-particle character of the ground
and lowest excited states of $^{57}$Ni. The spectroscopic factors
for the first three states of $^{57}$Ni were measured in a $(d,p)$
transfer reaction at low energy \cite{rehm98}.  In contrast
to the indications of collectivity from the measured $^{56}$Ni
$B(E2)$ excitation strength, the transfer
experiment indicated that 
the first three states of $^{57}$Ni are almost pure
single-particle states. 

Pickup and stripping reactions and extraction of 
spectroscopic factors can be used to resolve the nature of $^{56}$Ni and 
the single-particle character of $^{57}$Cu and $^{57}$Ni.
The accuracy of spectroscopic factors measured in this mass region
via low-energy transfer reactions is limited by the lack of good
optical model potential parameters. In experiments performed at
high energies, analyses based on the sudden approximation and eikonal
theory \cite{hans96,tost99} are applicable and thus the 
model dependency is reduced. The corresponding experimental
technique is one-nucleon knockout in inverse kinematics, a method
which has been used to measure single-particle configurations over
a range of nuclei since its development as a spectroscopic tool
\cite{hans01,hans03,gad04,ter04,gad04b}. In this method,
single-particle spectroscopy on the 
nucleus of interest is performed by extracting the following
quantities: partial cross sections to ground and
excited states of the knockout residues; the spectroscopic factors
for the removal of a nucleon from a specific single-particle orbit of the
projectile; and the orbital angular momentum of the removed nucleon.
Measurements are performed in inverse kinematics at beam energies
of greater than 50~MeV/nucleon.

In previous one-nucleon knockout experiments on well-bound nuclei, a
reduction in measured spectroscopic strength with respect to
shell-model predictions has been observed.
The resulting reduction factor $R_s$, the ratio of experimental and theoretical
spectroscopic strength, is 0.5-0.7 for one-proton and one-neutron
knockout from 
stable $^{12}$C and $^{16}$O 
\cite{brow02} and for one-neutron removal from well-bound $N=16$
isotones \cite{gad04}, 
but closer to unity for one-proton knockout from the lighter weakly
proton-bound nuclei 
$^{8}$B, $^{9}$C \cite{ende03} and for the one-neutron knockout from
the weakly neutron-bound $^{15}$C to the ground state of $^{14}$C
\cite{ter04}. Recently, a pronounced reduction of 0.24(4) has been
observed in the one-neutron removal from $^{32}$Ar \cite{gad04b}, the
most deeply-bound neutron system studied so far. 

\section{Theoretical analysis\label{theory}}

In the technique of one-nucleon knockout, the measured cross
sections are used to derive the one-nucleon removal
spectroscopic factors via an extension of the eikonal model
\cite{tost99}. The cross section $\sigma$ to a specific final state
of the knockout residue (core) is related to the spectroscopic factor via
\begin{eqnarray}
\label{spec_eq}
\sigma_{exp} &=&\sum_{j} C^{2}S_{exp}(nlj)
\, \sigma_{{\rm sp}}(S_{n},nlj),\\
\sigma_{th} &=&\sum_{j} \left( \frac{A}{A-1} \right) ^{N} C^{2}S_{SM}(nlj)
\, \sigma_{{\rm sp}}(S_{n},nlj),
\end{eqnarray}
where the sum is taken over all non-vanishing nucleon-core
configurations. The quantity $C^{2}S$ is the spectroscopic factor
for the removal of a nucleon with given single-particle quantum
numbers $nlj$. The factor $(A/(A-1))^{N}$ is a center-of-mass
correction that has to be applied if spectroscopic factors are taken
from shell model to obtain a theoretical cross section. In the
center-of-mass correction, $A$ is the mass 
number of the initial nucleus and $N$ is the main harmonic
oscillator quantum number associated with the relevant major shell
\cite{brow02}. The spectroscopic factor depends on the structure
of the nucleus, while the single-particle cross section
$\sigma_{{\rm sp}}$ comes from reaction theory, assuming unit
single-particle strength.

The single-particle cross sections are highly dependent on the
nucleon separation energy $S_{\rm n,p}$ and the orbital angular momentum $l$
of the removed nucleon. These were calculated in the eikonal model
and include contributions from both the stripping and diffraction
processes \cite{tost99}. The removed-neutron wave functions in the
$^{57}$Ni ground state were calculated in Woods-Saxon potential
wells with a diffuseness 0.7~fm. Their radius parameters were
chosen so that the {\it rms} radius of each neutron orbital was 
consistent with those from a Skyrme SKX Hartree-Fock (HF) \cite{bro98}
calculation for
$^{57}$Ni when the depth of the Woods-Saxon potential was adjusted to
reproduce the effective separation energy. The measured separation
energy of 
$S_{\rm n}=10.247$~MeV \cite{bhat98} was used for the $p_{3/2}$
state and an effective separation energy of 14~MeV was estimated from
the centroid of the  
$p_{3/2}f_{7/2}$ multiplet for the $f_{7/2}$ level. The
orbital {\it rms} radii were assumed to be 4.14 and 4.17~fm  for the
$p_{3/2}$ 
and $f_{7/2}$ levels. The $^{56}$Ni neutron and proton
densities were also taken from HF, resulting in a {\it
rms} matter radius of 3.64~fm for $^{56}$Ni, consistent with $^{56}$Fe. A
Gaussian matter distribution was assumed for the $^9$Be target
with an {\it rms} matter radius of 2.36~fm.

Excited levels of $^{56}$Ni and associated spectroscopic factors
were calculated in the many-body shell model \cite{brow88,warb92}
using the FPD6 effective interaction \cite{semo96}. For
$^{57}$Ni, the $\nu(f_{7/2})$ and $\pi(f_{7/2})$ closed shells
plus one neutron in the $p_{3/2}$ orbit were assumed. For
$^{56}$Ni, calculations were performed with the $\nu(f_{7/2})$ and
$\pi(f_{7/2})$ closed shells for the ground state, and for the
excited states $\pi(f_{7/2})^{8}$ and $\nu(f_{7/2})^{7}$ with the
$p_{3/2}$, $f_{5/2}$, and $p_{1/2}$ neutron shells active. 
In this approach, the spectroscopic factor for the removal of a neutron from
the  $p_{3/2}$ orbit leading to the ground
state of $^{56}$Ni is
$C^2S_{SM}=1$ and the spectroscopic factors for removing a neutron
from $f_{7/2}$ summed over all 
final excited
states of $^{56}$Ni ($T=0$ and $T=1$) is $C^2S_{SM}=8$. Considering
the calculated energy 
levels of $^{56}$Ni, we
predict 7.36 of this to be below the proton separation energy.

The shape of the measured parallel momentum distribution of the
$^{56}$Ni residues is used to assign the orbital angular momentum
for the removed nucleon. The theoretical parallel momentum
distributions were calculated in a black-disk model 
\cite{hans96}, core-target collisions at small impact parameters
($<$7.2~fm) being assumed to result in fragmentation of the core.
Interaction radii were chosen to reproduce the reaction cross
sections of the free constituents. 

\section{Experiment and data analysis\label{exptsetup_sec}}

\begin{figure}
\includegraphics[width=8cm]{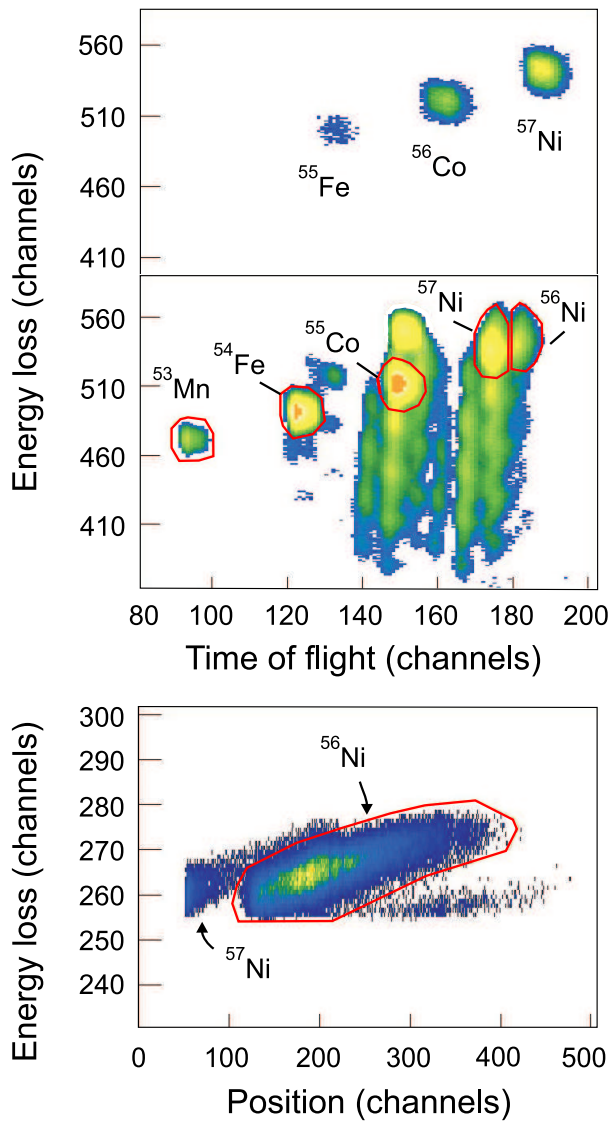}
\caption{(Color online) Particle identification in the S800 focal
  plane. The upper 
  panel shows the energy loss measured in the ionization chamber versus the
  time of flight taken between two plastic scintillators for the
  unreacted beam (top) and the setting with the spectrograph centered on
  the one-neutron removal (bottom). The lower
  panel shows the correlation between the energy loss and the
  x-position (dispersive) in the focal plane. In this spectrum, a
  beam blocker had been 
  inserted at the high-momentum side to block the tail of the
  unreacted  $^{57}$Ni beam.  $^{56}$Ni and $^{57}$Ni
  isotopes can be clearly separated. The x-position relates to the parallel
  momentum. \label{pid}}  
\end{figure}

The experiment was performed at the National
Superconducting Cyclotron Laboratory (NSCL). Acceleration of a
beam of stable $^{58}$Ni to an energy of 105~MeV/nucleon and an
intensity of 2~pnA was carried out in the K1200 cyclotron of the
NSCL. The $^{58}$Ni beam was incident on a 249~mg/cm$^{2}$
$^{9}$Be production target. The A1200 fragment separator
\cite{sher92} was used to select a secondary beam of $^{57}$Ni
with an energy of 73~MeV/nucleon, an intensity of approximately
30,000 particles per second and a momentum spread of 0.5\%. A
56.1~mg/cm$^{2}$ $^{9}$Be target was located at the center of an
array of 38 position-sensitive NaI(Tl) detectors \cite{sche99} at the
entrance to the 
S800 spectrograph \cite{baz03}. Gamma rays were measured in
coincidence with the knockout residues detected by the S800
focal-plane detector system \cite{yurk99}. The identification of the 
$^{56}$Ni fragments was performed with energy loss and position
information  
measured with the S800 spectrograph focal-plane detectors and using the
time of flight taken between plastic scintillators at the
exit of the A1200 fragment separator and the S800 focal plane (see
Fig.\ref{pid}). 


\begin{figure}
\includegraphics[width=3.375in]{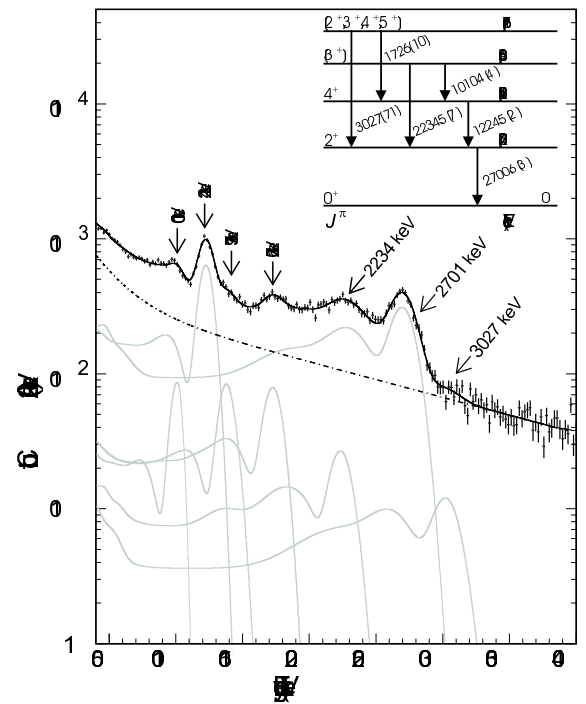}
\caption{Doppler-reconstructed ($\beta=0.36$) projectile-frame
$\gamma$-ray spectrum in coincidence with $^{56}$Ni fragments detected
in the spectrograph focal plane. The solid black line is the fit to
the experimental spectrum containing the sum of the simulated response
functions for seven $\gamma$-ray peaks (solid grey lines) and a
double-exponential 
coincident background (dashed line), assumed to arise from excitations
of the target. The inset shows a partial decay scheme for $^{56}$Ni
with a new proposed level at 5661(72) keV. 
\label{gamspec}}
\end{figure}

After the $^{56}$Ni residues from one-neutron knockout were
unambiguously identified, the coincident $\gamma$-ray spectrum was
analyzed. Only the inner ring of 11 NaI(Tl) detectors
\cite{sche99} was used in this experiment. GEANT \cite{geant}
simulations successfully modeled the $\gamma$-ray spectrum and the
detector response. Simulations and experimental efficiencies measured with
standard calibration sources agreed within 7.5\%.
The Doppler-reconstructed ($\beta=0.36$) $\gamma$-ray spectrum in
coincidence with $^{56}$Ni fragments is shown in
Fig.~\ref{gamspec}. The analytical curves, 
which were fitted 
to the simulated spectra and subsequently scaled to fit the
experimental spectrum, can be seen as solid grey lines. The dashed
line indicates the prompt, coincident background, described in
this experiment by a double exponential curve. A similar
background has been observed in previous knockout experiments
\cite{auma00,madd00,madd01} where it has been attributed to neutrons,
$\gamma$ rays and charged particles interacting with the experimental
apparatus and scintillators. The background determined for this
experiment was 67\% to 85\% higher than that quoted for knockout
on Si and S \cite{ende02}. There may be evidence for an increase
in coincident background with mass number or binding energy of the
knocked-out nucleons.

\section{Results and discussion}

A simplified level scheme including all $\gamma$-ray transitions
observed in this experiment is shown in the inset of
Fig.~\ref{gamspec}. The placement of the 1726(10) and
3027(71)~keV $\gamma$ rays in coincidence with the 1224.5(7) and
2700.6(3)~keV $\gamma$ rays resulted in a level at 5661(72)~keV. A
5668~keV level and associated 1744~keV de-excitation $\gamma$ ray
were observed in 1985 \cite{blom85} and reported without uncertainty.
While an angular momentum assignment of \mbox{$J^{\pi}=6^{+}$} was
proposed for the 5668~keV level, only
\mbox{$J^{\pi}=2^{+},3^{+},4^{+},5^{+},$} are expected to be populated via
the one-neutron knockout mechanism due to angular momentum considerations. No
placement in the proposed level scheme was possible for the $\gamma$
ray observed at 1379(10)~keV.

The proton separation energy for $^{56}$Ni \mbox{($S_{\rm
p}=7.165(11)$~MeV)} is significantly higher than the energy of the
highest excited state observed in this experiment. While excited
states up to an energy of 7900~keV were predicted in a shell-model
calculation, no higher-lying excited states could be identified
experimentally above 5661~keV. Low statistics at high $\gamma$-ray energies
contributed to the difficulty in resolving any transitions above
3100~keV. Therefore the branching ratios to the individual excited
states of $^{56}$Ni were not calculated due to the possibility of
indirect feeding from higher-lying, unobserved excited states.


\begin{figure}
\includegraphics[width=3.375in]{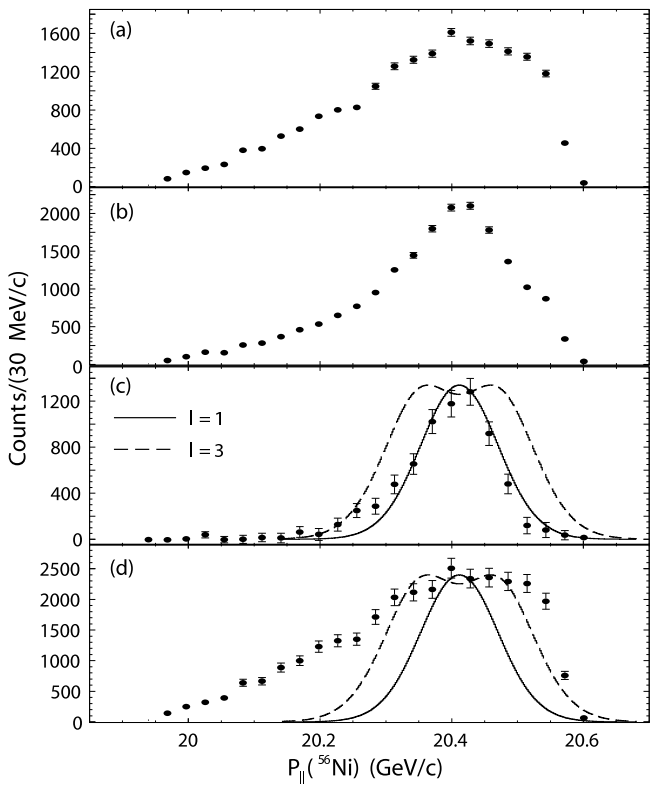}
\caption{Parallel momentum distributions associated with $^{56}$Ni
detected at the S800 focal plane. The distribution
associated with $^{56}$Ni fragments in coincidence with all $\gamma$
rays between 250 and 7000~keV is shown in panel (a), and that
associated with all remaining $^{56}$Ni fragments in panel (b). The
distribution associated with knockout to all excited states of
$^{56}$Ni is shown in panel (d) and that to the ground state in
panel (c). The distributions in panels (c) and (d) were constructed
assuming an average efficiency for the NaI(Tl) array of
\mbox{$\epsilon=0.55$} and a background probability of
\mbox{$\delta = 0.1$} (see text for details). \label{bothmom}}
\end{figure}

The parallel momentum distributions of the $^{56}$Ni knockout
residues were determined using the reconstructed scattering angle
and fractional kinetic energy after the
target, the momentum (20.26~GeV/$c$) and beam velocity
$\beta=0.36$ at the point of the $\gamma$-ray emission, and the mass of
the fragments. Due to the complexity of the $\gamma$-ray spectrum
of $^{56}$Ni, it was not possible to
isolate momentum distributions associated with individual final
excited states. Instead, the momentum distributions associated
with the knockout to all excited states and the ground state of $^{56}$Ni were
reconstructed from the measured momenta in coincidence and
anti-coincidence with detected $\gamma$ rays. The coincident ($C$)
and anti-coincident ($A$) spectra shown in Fig.~\ref{bothmom}(a)
and (b), respectively, can be written in terms of the momentum
distributions to the ground state ($S_{\rm g.s.}$) and all excited
states ($S_{\rm exc}$) of $^{56}$Ni as
\begin{equation}
\begin{array}{c}
S_{\rm exc}(p_{\parallel}) = \frac{1}{\epsilon} \left( C(p_{\parallel})
- \frac{\delta}{1-\delta} A(p_{\parallel}) \right)\\
\\
S_{\rm g.s.}(p_{\parallel}) = \left( 1 + \frac{\delta}{\epsilon(1-
\delta)} \right)
\left( A(p_{\parallel}) - \frac{(1-\delta)(1-\epsilon)}{\epsilon +
\delta - \epsilon \delta} C(p_{\parallel}) \right).
\end{array}
\end{equation}
The average efficiency $\epsilon$ for the NaI(Tl) array is
expected to be approximately 50\% for an average $\gamma$-ray cascade
of two, but was varied in the analysis to determine the best
separation between $S_{\rm g.s.}$ and $S_{\rm exc}$. The
probability that a $\gamma$ ray not originating from the
de-excitation of a knockout residue would be detected in
coincidence with that fragment in the S800 focal plane is
$\delta$. For this experiment, a value of \mbox{$\delta = 0.1$}
was chosen, in agreement with previous experiments which
determined a $\gamma$-ray background of 10\% per fragment. The
background appears to be higher in this experiment than the 10\%
assumed in experiments on lighter nuclei. However, with this
choice of $\delta$, a consistent separation of the ground state
and excited state events was obtained with a choice of
\mbox{$\epsilon = 0.55$}. Thus the higher background used in the
fit to the $\gamma$-ray spectrum in this experiment may be a result
of multiple $\gamma$ rays that could not be resolved due to the level
density and low statistics at high energies. A 3.5\% systematic
error on the number of counts was included in the calculation of
experimental cross sections due to the choice of the two
parameters $\epsilon$ and $\delta$. The resulting momentum
distributions associated with 
the ground state and all excited states of $^{56}$Ni are shown in
Fig.~\ref{bothmom}(c) and (d). The errors shown include both, 
statistical uncertainties and the systematic error from the choice of
$\epsilon$ and $\delta$. A similar analysis has been performed for the
one-nucleon removal reactions on $sd$ shell nuclei~\cite{Nav98}.

The shell model predicts that one-neutron removal to the ground
state would be associated with an \mbox{$l=1$} distribution, due
to the removal of the $1p_{3/2}$ valence neutron. Similarly,
one-neutron removal to the excited states of $^{56}$Ni would
correspond to an \mbox{$l=3$} momentum distribution due to the
orbital angular momentum carried by a neutron removed from the
$0f_{7/2}$ shell. 
The ground state distribution shown in Fig.~\ref{bothmom}(c) is
well described by an \mbox{$l=1$} theoretical curve centered at
20.42 GeV/$c$. The distribution associated with knockout to the
excited states of the $^{56}$Ni fragments, shown in
Fig.~\ref{bothmom}(d) is consistent with the theoretical curve for
\mbox{$l=3$} with a low-momentum excess extending to approximately
20~GeV/$c$. This low-momentum tail as seen around 20.25 GeV/$c$ in the
excited-state momentum distribution (d), has been reported in several previous
one-nucleon knockout experiments \cite{auma00,madd01c,ende02,gad05}.
Tostevin \cite{tost01} has shown that the asymmetric shape as
observed for the knockout 
reactions on the halo nuclei $^{11}$Be and $^{15}$C can be
reproduced by using a proper dynamical treatment of continuum-coupling
effects in the diffractive channel of the knockout process. However,
the same explanation is 
unlikely to hold for the well-bound $^{57}$Ni. The present observation
is very similar to the pronounced tail at low momentum 
reported in the one-neutron knockout from $^{46}$Ar to the $7/2^-$
ground state of $^{45}$Ar \cite{gad05}, where this asymmetry is
discussed in the framework of deviations from eikonal theory.


\begin{table}
\caption{
Experimental cross sections $(\sigma _{exp})$ in mb and spectroscopic
factors (C$^{2}$S$_{exp}=\sigma _{exp}$/$\sigma _{sp})$ for the
various final states of $^{56}$Ni populated in the
$^{9}$Be($^{57}$Ni,$^{56}$Ni+$\gamma )$X reaction at 73 MeV/nucleon.
Theoretical single-particle cross sections ($\sigma _{sp}=\sigma
_{str}+\sigma _{diff})$ were calculated in the eikonal model
\cite{hans96,tost99} for
the stripping $(\sigma _{\rm str})$ and diffractive $(\sigma _{\rm diff})$
processes. Spectroscopic factors from shell model (C$^{2}$S$_{SM}$)
and the theoretical cross section $\sigma_{\rm th}$, which combines the
spectroscopic factors from the shell model and the single-particle cross
section from the reaction theory following eq.~(2), are
compared to the experimental values in terms of the ratio $R_{s}=\sigma
_{exp}$/$\sigma _{th}$ (see text).\label{xsec}}
\begin{ruledtabular}
\begin{tabular}{llll}
& ground state & excited states & inclusive \\
\hline
I$^{\pi}$ & 0$^{+}$ & 2$^{+}$--5$^{+}$ & \\
$l$ & 1 & 3 & \\
\hline
$\sigma_{\rm str}$\small{(mb)}   & 9.86 & 7.2 & \\
$\sigma_{\mathrm{diff}}$\small{(mb)}& 3.48 & 2.0 & \\
$\sigma_{\rm sp}$\small{(mb)}    & 13.3 & 9.2 & \\
\hline
$\sigma_{\rm exp}$\small{(mb)}&  7.7(15)& $33.7(17)$ & $41.4(12)$ \\
$C^{2}S_{\rm exp}$ &        $0.58(11)$ & $3.7(2)$ & \\
\hline
$C^{2}S_{\rm SM}$          & 1.0 & 7.36 & \\
$\sigma_{\rm th}$\small{(mb)}  & 14.0 & 71.4 & 85.4 \\
\hline

$R_{s}$   & $0.55(11)$ & $0.47(2)$ & $0.48(2)$ \\
\end{tabular}
\end{ruledtabular}
\end{table}
The inclusive cross section $\sigma_{\rm incl}$ was calculated
from the number of $^{56}$Ni fragments detected in the spectrograph
focal plane relative to the number of incident $^{57}$Ni projectiles,
normalized 
to the number of target nuclei. The
inclusive cross section was calculated run by run, and the average
for the three runs closest to the unreacted $^{57}$Ni normalization
run was used
to determine $\sigma_{\rm incl}=41.4(12)$~mb. The partial cross
sections to all excited states and the ground state of $^{56}$Ni
were determined using the same method as for the parallel momentum
distributions. This resulted in branches of 81.3(35)\% to all
excited states and 18.7(35)\% to the ground state, corresponding
to a partial cross section to the ground state of 7.7(15)~mb and
to all excited states of 33.7(17)~mb. The calculated stripping and
diffraction components for the single-particle cross sections to
the ground state and excited states of the fragments are listed in
Table~\ref{xsec}, along with the total theoretical cross sections
to both final states. The deduced spectroscopic factors were
\mbox{$C^{2}S_{\rm exp}=0.58(11)$} for the ground state and
\mbox{$C^{2}S_{\rm exp}=3.7(2)$} for all excited states.


\begin{figure}
\includegraphics[width=3.375in]{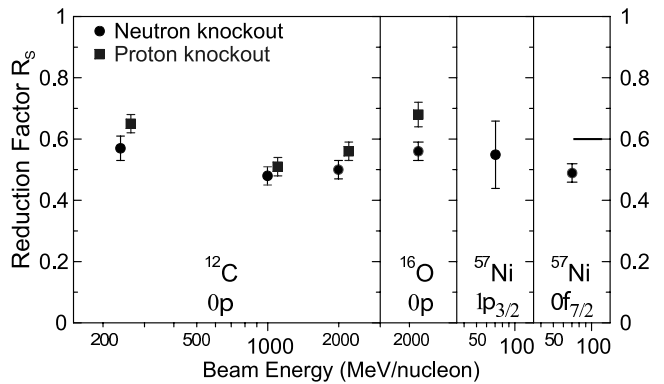}
\caption{Reduction factor $R_{s}$ versus incident beam energy
for one-nucleon knockout from the well-bound nuclei $^{12}$C,
$^{16}$O and $^{57}$Ni. The circles
(squares) indicate results from one-neutron (proton) knockout.
$R_{s}$ values for $^{12}$C and $^{16}$O were calculated in
\cite{brow02} from measurements performed at the Lawrence Berkeley National
Laboratory \cite{kidd88,olse83}.
\label{quenchfig}}
\end{figure}

The shell model is expected to most accurately predict the
properties of nuclei at or near closed shells. As expected for
$^{57}$Ni, a semi-magic nucleus with only one neutron outside the
doubly-magic $N=Z=28$ core, the shell model correctly predicts the
angular momentum of the knocked-out neutron for the ground-state
configuration, shown by the good fit of the \mbox{$l=1$} shape to the
momentum distribution associated with the $^{56}$Ni ground
state. Although the orbital angular momentum 
of knocked-out neutrons populating excited states of $^{56}$Ni
could not be well 
determined, a value of at least \mbox{$l=3$} should be
assigned. The measured partial and inclusive cross sections,
however, do not show good agreement with theory. The measured
cross section to the ground state is 55(11)\% of the theoretical
ground-state cross section. The measured cross section to all
excited states amounts to 47(2)\% of the shell-model sum-rule theoretical cross
section, and the inclusive cross section is 48(2)\% of the
theoretical inclusive cross section. The reduction in cross section
translates into a reduction in measured spectroscopic 
strength relative to the shell model.

Several studies have been performed recently to explore this
reduction in spectroscopic strength observed in one-nucleon
removal experiments \cite{brow02,ende03,ter04,gad04b,gad04,gad05}. 
Proton removal from inelastic electron scattering has been considered
the standard for the determination of absolute spectroscopic
factors. It was  
shown that deduced spectroscopic factors from ($e,e'p$)
experiments for nuclei from mass 12 to 208 exhaust only about 60\%
of the single-particle shell model prediction \cite{pand97,kram01,dick04}.
Electron scattering data is only available for stable nuclei, and
it is not possible to probe neutron occupancies within this
approach. In contrast, the technique of one-nucleon 
knockout can be used to probe both proton and neutron occupancies
in radioactive nuclei as well as stable species. It then becomes of
interest to compare the spectroscopic factors deduced via
one-proton knockout reactions to those from electron scattering. Brown
{\it et al.\ }\cite{brow02} compared 
theoretical cross sections calculated in the eikonal model with
experimental cross sections from a series of one-nucleon knockout
experiments. In order to facilitate a comparison between
experiments, the reduction factor $R_{s}$ was defined as the ratio
between the experimental and theoretical cross sections.
One-proton knockout on the stable nuclei $^{12}$C and $^{16}$O
resulted in $R_{s}$ values of 0.53(2) and 0.68(4), in agreement
with the reduction factors deduced from the electron scattering
experiments. One-neutron knockout from the same two stable nuclei
yielded $R_{s}$ values of 0.49(2) and 0.56(3), suggesting that
one-neutron knockout experiments measure the same quantity as
one-proton knockout, and thus can also be used to measure absolute
spectroscopic factors. The reduction in single-particle strength
observed in the case of the radioactive nucleus $^{57}$Ni is
comparable to the reductions obtained for the single-neutron knockout of
the two stable nuclei $^{12}$C and $^{16}$O \cite{brow02} (see
Fig.~\ref{quenchfig}), consistent with the reduction observed in the
one-neutron knockout on well-bound $N=16$ isotones \cite{gad04} and
agrees with the quenching of single-particle strengths observed for 
stable nuclei in electron scattering experiments \cite{pand97,kram01,dick04}.

It is suggested \cite{pand97,dick04,ende03,gad04b} that correlation
effects (short-range, 
long-range and tensor nucleon-nucleon interactions) which are absent
from or only
approximately treated in effective-interaction theory may
explain some of the observed reduction in spectroscopic strength. Those
correlations result in physical nucleon occupancies being
reduced and spectroscopic strength shifted to higher energies; see
\cite{pand97,dick04}. For example, the repulsive short-range part of
the interaction becomes important at distances of less than 0.4\ fm,
and leads to high-momentum components in the nucleon wave function
\cite{pand97}. The importance of the 
coupling to surface phonons and giant resonances as contribution to
the reduction has been discussed in \cite{dick04}.  

In summary, spectroscopic factors for one-neutron removal from $^{57}$Ni to
the ground and all observed excited states of $^{56}$Ni have been
measured at 73~MeV/nucleon. The $l=1$
character of the $^{57}$Ni ground state and the $l=3$ character of
the first few excited states have been confirmed. The level scheme
of $^{56}$Ni was extended to include a level at 5661(72)~keV,
with two de-excitation $\gamma$ rays. Finally, the phenomenon of a
reduction in the measured spectroscopic strengths, as compared to
shell-model predictions, has been confirmed in the exotic, doubly-magic
$^{56}$Ni -- the heaviest nucleus yet measured 
via one-nucleon knockout. The observed inclusive cross section for
knockout to $^{56}$Ni was 48(2)\% of the predicted shell-model
value, in agreement with the reductions typically observed in
stable, well-bound nuclei.

\section*{Acknowledgements}
This work was supported by the National Science Foundation under
grants PHY-0110253 and PHY-0244453, and by the United
Kingdom Engineering and Physical Sciences Research Council (EPSRC) under
grant EP/D003628.
.

\end{document}